\documentclass[aps,twocolumn,superscriptaddress,prl,floatfix]{revtex4}
\usepackage{amssymb}
\usepackage{amsmath}
\usepackage{graphicx}

\def\gmat{\mathbf{g}}
\def\E{\mathbf{E}}
\def\ed{\varepsilon_{d}}
\def\Ohm{\Omega}
\def\kb{k_{\mathrm{B}}}
\def\te{T_{\mathrm{e}}}

\def\Sione{S_{\mathrm{1}}}
\def\Sitwo{S_{\mathrm{2}}}

\def\Svx{S_{\mathrm{V12}}}
\def\Svonetwo{S_{\mathrm{V1,2}}}
\def\EP{S^{\mathrm{EP}}}

\def\EPone{S_{\mathrm{1}}^{\mathrm{EP}}}

\def\SPone{S^{\mathrm{P}}_{\mathrm{1}}}
\def\Six{S_{\mathrm{12}}}
\def\F{\mathcal{F}}
\def\vtl{V_{\mathrm{tl}}}

\def\vbl{V_{\mathrm{bl}}}
\def\vbc{V_{\mathrm{bc}}}
\def\vbr{V_{\mathrm{br}}}
\def\vl{V_{\mathrm{l}}}
\def\vr{V_{\mathrm{r}}}

\def\Ione{I_{\mathrm{1}}}
\def\Itwo{I_{\mathrm{2}}}

\def\vzero{V_{\mathrm{0}}}
\def\vone{V_{\mathrm{1}}}

\def\gzeroone{g_{\mathrm{01}}}

\def\gonetwo{g_{\mathrm{12}}}
\def\gtwoone{g_{\mathrm{21}}}

\def\MHz{\mathrm{MHz}}

\def\Hethree{^3\mathrm{He}}

\begin{document}
\title{Noise Correlations in a Coulomb Blockaded Quantum Dot}
\author{Y.~Zhang}
\affiliation{Department of Physics, Harvard University, Cambridge,
Massachusetts 02138, USA}
\author{L.~DiCarlo}
\affiliation{Department of Physics, Harvard University, Cambridge,
Massachusetts 02138, USA}
\author{D.~T.~McClure}
\affiliation{Department of Physics, Harvard University, Cambridge,
Massachusetts 02138, USA}
\author{M.~Yamamoto}
\affiliation{Department of Applied Physics, University of Tokyo,
Bunkyoku, Tokyo 113-8656, Japan} \affiliation{SORST-JST,
Kawaguchi-shi, Saitama 331-0012, Japan}
\author{S.~Tarucha}
\affiliation{Department of Applied Physics, University of Tokyo,
Bunkyoku, Tokyo 113-8656, Japan}  \affiliation{ICORP-JST,
Atsugi-shi, Kanagawa 243-0198, Japan}
\author{C.~M.~Marcus}
\affiliation{Department of Physics, Harvard University, Cambridge,
Massachusetts 02138, USA}
\author{M.~P.~Hanson}
\affiliation{Department of Materials, University of California,
Santa Barbara, California, 93106, USA}
\author{A.~C.~Gossard}
\affiliation{Department of Materials, University of California,
Santa Barbara, California, 93106, USA}
\date{\today}

\begin{abstract}
We report measurements of current noise auto- and cross-correlation
in a tunable quantum dot with two or three leads. As the Coulomb
blockade is lifted at finite source-drain bias, the auto-correlation
evolves from super-Poissonian to sub-Poissonian in the two-lead
case, and the cross-correlation evolves from positive to negative in
the three-lead case, consistent with transport through multiple
levels. Cross-correlations in the three-lead dot are found to be
proportional to the noise in excess of the Poissonian value in the
limit of weak output tunneling.
\end{abstract}

\maketitle

Considered individually, Coulomb repulsion and Fermi statistics both
tend to smooth electron flow, thereby reducing shot noise below the
uncorrelated Poissonian limit~\cite{blanter00-05,buttiker90-92}. For
similar reasons, Fermi statistics without interactions also induces
a negative noise cross-correlation in multiterminal
devices~\cite{blanter00-05,buttiker90-92,henny99,oliver99}. It is
therefore surprising that under certain conditions, the interplay
between Fermi statistics and Coulomb interaction can lead to
electron bunching, i.e., super-Poissonian auto-correlation and
positive cross-correlation of electronic noise.

The specific conditions under which such positive noise correlations
can arise has been the subject of numerous
theoretical~\cite{Iannaccone98,texier00,buttiker03,wu05,rychkov06,sukhorukov01,kiesslich03,belzig05,cottet04,thielmann05}
and
experimental~\cite{Iannaccone98,safonov03,barthold06,chenPRB06,chenPRL06,oberholzer06,onac06,gustavsson06,Zarchin06,McClure06}
studies in the past few years. Super-Poissonian noise observed in
MESFETs~\cite{safonov03}, tunnel barriers~\cite{chenPRB06} and
self-assembled stacked quantum dots~\cite{barthold06} has been
attributed to interacting localized
states~\cite{eto97,kiesslich03,safonov03} occurring naturally in
these devices.  In more controlled geometries, super-Poissonian
noise has been associated with inelastic
cotunneling~\cite{sukhorukov01} in a nanotube quantum
dot~\cite{onac06}, and with dynamical channel
blockade~\cite{belzig05,cottet04} in  GaAs/AlGaAs quantum dots in
the weak-tunneling~\cite{gustavsson06} and quantum Hall
regimes~\cite{Zarchin06}. Positive noise cross-correlation has been
observed in a capacitively-coupled double dot~\cite{McClure06} as
well as in electronic beam-splitters following either an inelastic
voltage probe~\cite{oberholzer06,texier00,buttiker03,wu05,rychkov06}
or a super-Poissonian noise source~\cite{chenPRL06}. The predicted
positive noise cross-correlation in a three-lead quantum
dot~\cite{cottet04} has not been reported experimentally to our
knowledge.

\begin{figure}[b!]
\center \label{figure1}
\includegraphics[width=2.95in]{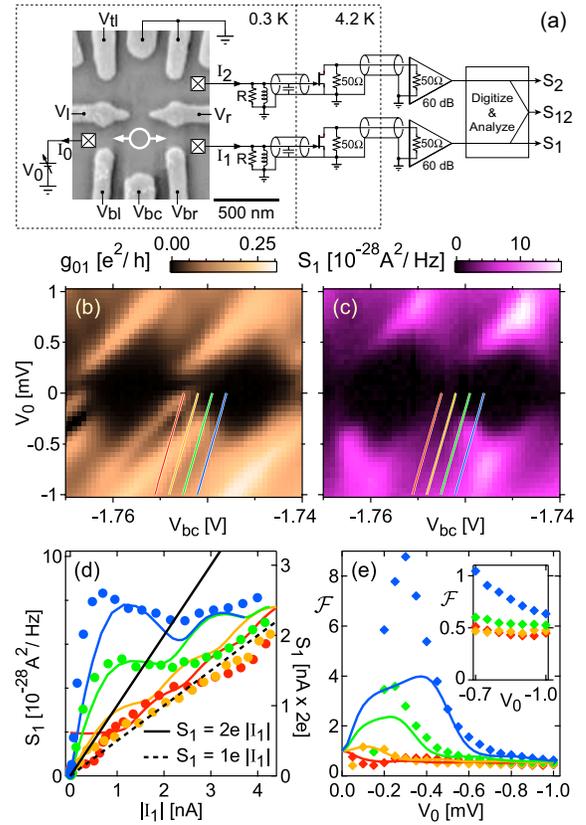}
\caption{\footnotesize{(a) Micrograph of the device and equivalent
circuit near $2~\MHz$ of the noise detection system (see text for
equivalent circuit near dc). For the data in Figs.~1 and 2, the
$\vl$-$\vr$ constriction  is closed and the dot is connected only to
reservoirs~0 and 1. (b, c) Differential conductance $\gzeroone$ and
current noise spectral density $\Sione$, respectively, as a function
of $\vzero$ and $\vbc$. (d) $\Sione$ versus $|\Ione|$ data (circles)
and multi-level simulation (solid curves) along the four cuts
indicated in (b) and (c) with corresponding colors. Black solid
(dashed) line indicates $\Sione = 2 e |\Ione|$ ($\Sione = 1e
|\Ione|)$. (e) Data (diamonds) and multi-level simulation (solid
curves) of the modified Fano factor $\F$ along the same cuts as
taken in (d). Inset: detail of $\F$ at high $|\vzero|$.}}
\end{figure}

This Letter describes measurement of current noise auto- and
cross-correlation in a Coulomb-blockaded quantum dot configured to
have either two or three leads. As a function of gate voltage and
bias, regions of super- and sub-Poissonian noise, as well as
positive and negative noise cross-correlation, are identified.
Results are in good agreement with a multi-level
sequential-tunneling model in which electron bunching arises from
dynamical channel blockade~\cite{belzig05,cottet04}. For
weak-tunneling output leads, noise cross-correlation in the
three-lead configuration is found to be proportional to the
deviation of the auto-correlation from the Poissonian value (either
positive or negative) similar to the relation found in electronic
Hanbury Brown--Twiss (HBT)--type
experiments~\cite{henny99,oliver99,chenPRL06}.

The quantum dot is defined by gates on the surface of a
$\mathrm{GaAs/Al_{0.3}Ga_{0.7}As}$ heterostructure [Fig.~1(a)]. The
two-dimensional electron gas $100~\mathrm{nm}$ below the surface has
density $2\times10^{11}~\mathrm{cm}^{-2}$ and mobility $2\times10^5~
\mathrm{cm}^2/\mathrm{Vs}$. Leads formed by gate pairs $\vl$-$\vbl$,
$\vr$-$\vbr$, and $\vl$-$\vr$ connect the dot to three reservoirs
labeled 0, 1, and 2, respectively. Plunger gate voltage $\vbc$
controls the electron number in the dot, which we estimate to be
$\sim 100$. The constriction formed by $\vtl$-$\vl$ is closed.

A $\Hethree$ cryostat is configured to allow simultaneous
conductance measurement near dc and noise measurement near
$2~\mathrm{MHz}$~\cite{techniques}. For dc measurements, the three
reservoirs are each connected to a voltage amplifier, a current
source, and a resistor  to ground ($r=5~\mathrm{k\Ohm}$). The
resistor $r$ converts the current $I_{\mathrm{\alpha}}$ out of
reservoir~$\alpha$ to a voltage signal measured by the voltage
amplifier; it also converts the current from the current source to a
voltage excitation $V_{\mathrm{\alpha}}$ applied at
reservoir~$\alpha$. The nine raw differential conductance matrix
elements  $\tilde{g}_{\mathrm{\alpha\beta}} =
dI_{\mathrm{\beta}}/dV_{\mathrm{\alpha}}$ are measured
simultaneously with lock-in excitations of $20~\mathrm{\mu V_{rms}}$
at 44, 20 and 36~$\mathrm{Hz}$ on reservoirs~0, 1 and 2,
respectively. Subtracting $r$ from the matrix $\tilde{\gmat}$ yields
the intrinsic conductance matrix $\gmat = [\E + r
\tilde{\gmat}]^{-1}\cdot\tilde{\gmat}$, where $\E$ is the identity
matrix.  Ohmic contact resistances ($\sim 10^3~\Ohm$) are small
compared to dot resistances ($\gtrsim 10^5~\Ohm$), and are neglected
in the analysis. Values for the currents $I_{\mathrm{\alpha}}$ with
bias $\vzero$ applied to reservoir 0 are obtained by numerically
integrating $\tilde{g}_{\mathrm{0\alpha}}$.

Fluctuations in currents $\Ione$ and $\Itwo$ are extracted from
voltage fluctuations  around $2~\mathrm{MHz}$ across separate
resistor-inductor-capacitor (RLC) resonators [Fig.~1(a)]. Power
spectral densities $\Svonetwo$ and cross-spectral density $\Svx$ of
these voltage fluctuations~\cite{techniques} are averaged over
$20~\mathrm{s}$, except where noted. Following the calibration of
amplifier gains and electron temperature $\te$ using noise
thermometry~\cite{techniques}, the dot's intrinsic current noise
power spectral densities $S_{1,2}$ and cross spectral density $\Six$
are extracted by taking into account the feedback~\cite{wu05} and
thermal noise from the finite-impedance external
circuit~\cite{ExtractSi}.

Figure 1(b) shows conductance $\gzeroone$ as a function of $\vbc$
and $\vzero$ in a two-lead configuration, i.e., with the $\vl$-$\vr$
constriction closed. The characteristic Coulomb blockade (CB)
diamond structure yields a charging energy $E_{\mathrm{C}} =
0.8~\mathrm{meV}$ and lever-arm for the plunger gate
$\eta_{\mathrm{bc}} = \Delta\ed / (e\Delta\vbc) = 0.069$, where
$\ed$ is the dot energy. The diamond tilt
$\eta_{\mathrm{bc}}/(1/2-\eta_{\mathrm{0}})$ gives the lever-arm for
reservoir~0: $\eta_{\mathrm{0}} = \Delta\ed / (e\Delta\vzero) =
0.3$. As shown in Fig. 1(d), current noise $\Sione$ along selected
cuts close to the zero-bias CB peak (red, orange cuts) is below the
Poissonian value $2 e |\Ione|$ at all biases $|\Ione|$, while cuts
that pass inside the CB diamond (green, blue cuts) exceed $2 e
|\Ione|$ at low currents, then drop below $2 e |\Ione|$ at high
currents. At finite $\te$, the current noise $\SPone = 2 e \Ione
\coth(e\vzero/2\kb\te)$ of an ideal Poissonian noise source at bias
$\vzero$ may exceed $2 e |\Ione|$ due to the thermal (Johnson) noise
contribution~\cite{sukhorukov01}. Accordingly, we define a modified
Fano factor $\F \equiv \Sione / \SPone$. Figure~1(e) shows regions
of super-Poissonian noise, $\F>1$, when the green and blue cuts are
within the CB diamond. For all cuts, $\F$ approaches 1/2 at large
bias.

Current noise can also be identified as sub- or super-Poissonian
from the excess Poissonian noise $\EPone \equiv \Sione - \SPone$
being negative or positive, respectively. Unlike $\F$, $\EPone$ does
not have divergent error bars inside the CB diamond, where currents
vanish. As shown in Fig.~2(a), in regions where both $\Ione$ and
$\Sione$ vanish, $\EPone$ also vanishes. Far outside the CB
diamonds, $\EPone$ is negative, indicating sub-Poissonian noise.
However, $\EPone$ becomes positive along the diamond edges,
indicating super-Poissonian noise in these regions.

We next compare our experimental results to single-level and
multi-level sequential-tunneling models of CB transport. The
single-level model yields exact expressions for average current and
noise~\cite{blanter00-05,thielmann05,HansresThesis}: $\Ione =
(e/h)\int d\varepsilon
\gamma_{0}\gamma_{1}(f_{1}-f_{0})/[(\gamma_{1}+\gamma_{0})^2/4+(\varepsilon-\varepsilon_{d})^2]$,
$\Sione = (2e^2/h)\int d\varepsilon \{ \gamma_{0}^2\gamma_{1}^2\cdot
[f_{0}(1-f_{0})+f_{1}(1-f_{1})]+\gamma_{0}\gamma_{1}[(\gamma_{1}-\gamma_{0})^2/4+(\varepsilon
- \varepsilon_{d})^2] \cdot
[f_{0}(1-f_{1})+f_{1}(1-f_{0})]\}/[(\gamma_{1}+\gamma_{0})^2/4+(\varepsilon-\varepsilon_{d})^2]^2$,
where $\gamma_{0(1)}$ is the tunneling rate to reservoir~0(1) and
$f_{0(1)}$ is the Fermi function in reservoir~0(1). The dot energy
$\ed$ is controlled by gate and bias voltages: $\ed =
-e\vbc\eta_{\mathrm{bc}} - e\vzero\eta_{0} - e\vone\eta_{1} +
\mathrm{const.}$ For the multi-level sequential-tunneling model, a
master equation is used to calculate current and noise, following
Refs.~\cite{belzig05,cottet04,hershfield93}. To model transport, we
assume simple filling of orbital levels and consider transitions to
and from $N$-electron states that differ in the occupation of at
most $n$ levels above (indexed $1$ through $n$) and $m$ levels below
(indexed $-1$ through $-m$) the highest occupied level in the
$(N+1)$-electron ground state (level 0)~\cite{ExplainMLmodel}.

\begin{figure}[t]
\center \label{figure2}
\includegraphics[width=3.1in]{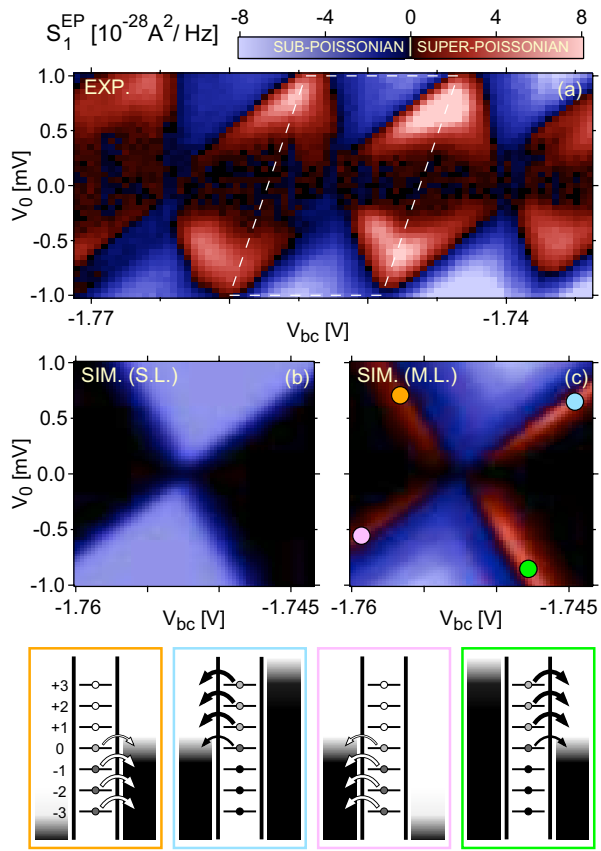}
\caption{\footnotesize{(a) Excess Poissonian noise $\EPone$ as a
function of $\vzero$ and $\vbc$. Red (blue) regions indicate
super(sub)-Poissonian noise. (b, c) Single-level (S.L.) and
multi-level (M.L.) simulation of $\EPone$, respectively,
corresponding to the data region enclosed by the white dashed
parallelogram in (a). At the four colored dots superimposed on (c),
where $\EPone$ is most positive, energy diagrams are illustrated in
the correspondingly colored frames at the bottom. In these diagrams,
black (white) arrows indicate electron (hole) transport; the
greyscale color in the reservoirs and inside the circles on each
level indicates electron population, the darker the higher.}}
\end{figure}

\begin{figure}[t]
\center \label{figure3}
\includegraphics[width=3.1in]{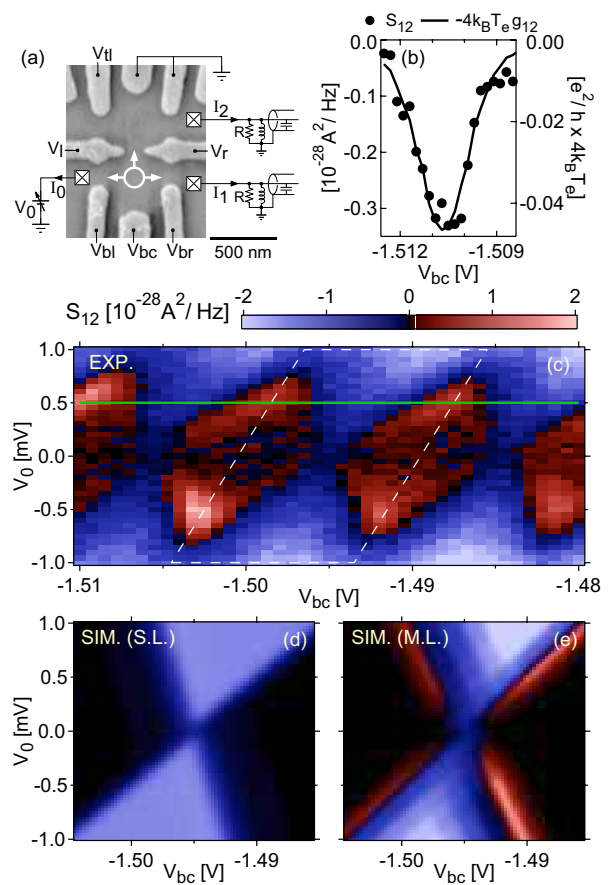}
\caption{\footnotesize{(a) The device in the three-lead
configuration, in which the data for this figure and for Fig.~4 are
taken. (b) $\Six$, integrated for $200~\mathrm{s}$, and
$-4\kb\te\gonetwo$ over a CB peak at zero bias. Left and right axes
are in different units but both apply to the data. (c) $\Six$ as a
function of $\vzero$ and $\vbc$. Red (blue) regions indicate
positive (negative) cross-correlation. (d, e) Single-level (S.L.)
and multi-level (M.L.) simulation of $\Six$, respectively,
corresponding to the data region enclosed by the white dashed
parallelogram in (c).}}
\end{figure}

Super-Poissonian noise in the multi-level model arises from
\textit{dynamical channel blockade}~\cite{belzig05,cottet04},
illustrated in the diagrams in Fig.~2. Consider, for example, the
energy levels and transport processes shown in the green-framed
diagram, which corresponds to the location of the green dot on the
lower-right edge in Fig.~2(c). Along that edge, the transport
involves transitions between the $N$-electron ground state and
$(N+1)$-electron ground or excited states. When an electron occupies
level~0, it will have a relatively long lifetime, as tunneling out
is suppressed by the finite electron occupation in reservoir~1 at
that energy. During this time, transport is blocked since the large
charging energy prevents more than one non-negative-indexed level
from being occupied at a time. This blockade happens dynamically
during transport, leading to electron bunching and thus to
super-Poissonian noise. At the location of the pink dot on the
lower-left edge in Fig.~2(c), the transport involves transitions
between the $(N+1)$-electron ground state and $N$-electron ground or
excited states; a similar dynamical blockade occurs in a
complementary hole transport picture. The hole transport through
level~0 is slowed down by the finite hole occupation in reservoir~0,
modulating the hole transport through negative-indexed levels, thus
leading to hole bunching and super-Poissonian noise. Transport at
the blue (orange) dot is similar to transport at the green (pink)
dot, but with the chemical potentials in reservoirs~0 and 1 swapped.
Both experimentally and in the multi-level simulation, $\EPone$ is
stronger along electron edges than along hole edges. This is due to
the energy dependence of the tunneling rates: since the
positive-indexed electron levels have higher tunneling rates than
the negative-index hole levels, the dynamical modulation due to
electron transport is stronger for electrons than for holes.

We next investigate the three-lead configuration, obtained by
opening lead~2 [Fig.~3(a)]. At zero bias, thermal noise
cross-correlation is found to be in good agreement with the
theoretical value~\cite{ExplainG12andG21}, $\Six = - 4 \kb \te
\gonetwo$, as seen in Fig.~3(b). To minimize this thermal
contribution to $\Six$, output leads are subsequently tuned to
weaker tunneling than the input lead ($g_{01} \sim g_{02} \sim
4g_{12}$), for reasons discussed below. Note that as a function of
$\vbc$ and $V_{0}$, $\Six$ [Fig.~3(c)] looks similar to $\EPone$
[Fig.~2(a)] in the two-lead configuration.

Both the single-level and multi-level models can be extended to
include the third lead~\cite{HansresThesis,cottet04}. Figures~3(d)
and 3(e) show the single-level and multi-level simulations of
$\Six$, respectively. Similar to the two-lead case, only the
multi-level model reproduces the positive cross-correlation along
the diamond edges.

\begin{figure}[t]
\center \label{figure4}
\includegraphics[width=3.25in]{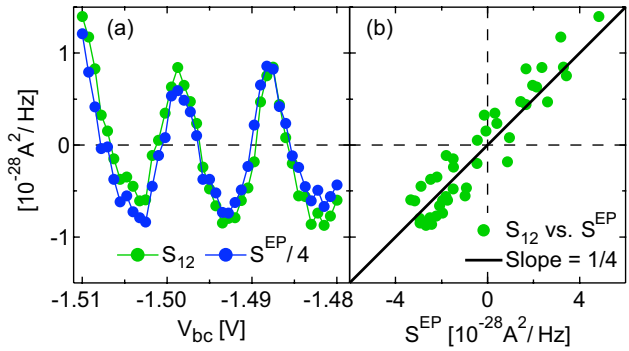}
\caption{\footnotesize{(a) $\Six$ (green) and $\EP / 4$ (blue) as a
function of $\vbc$ at $\vzero = +0.5~\mathrm{mV}$ (green horizontal
line in Fig. 3(c)). (b) Parametric plot of $\Six$ (green circles)
versus $\EP$ for the same data as in (a). The solid black line has a
slope of $1/4$, the value expected for a 50/50 beam-splitter.}}
\end{figure}

To further investigate the relationship between noise auto- and
cross-correlation, we compare $\Six$ to the total excess Poissonian
noise, $\EP \equiv \Sione + \Sitwo + 2 \Six - 2 e (\Ione + \Itwo)
\coth(e V_{0}/2\kb\te)$, measured in the same three-lead
configuration. Figure~4 shows $\EP$ and $\Six$, measured at fixed
bias $V_{0} = +0.5~\mathrm{mV}$. The observed proportionality $\Six
\sim \EP / 4$ is reminiscent of electronic HBT-type
experiments~\cite{henny99,oliver99,chenPRL06}, where noise
cross-correlation following a beam-splitter was found to be
proportional to the total output current noise in excess of the
Poissonian value, with a ratio of $1/4$ for a $50/50$ beam-splitter.
In simulation, we find that this HBT-like relationship holds in the
limit $g_{01} \sim g_{02} \gg g_{12}$ (recall that $g_{01} \sim
g_{02} \sim 4g_{12}$ in the experiment); on the other hand, when
$g_{01} \sim g_{02} \sim g_{12}$, thermal noise gives a negative
contribution that lowers $\Six$ below $\EP / 4$, as we have also
observed experimentally (not shown). The implications are that
first, with weak tunneling output leads, the three-lead dot behaves
as a two-lead dot followed by an ideal beam-splitter, and second,
the dynamical channel blockade that leads to super-Poissonian noise
in the two-lead dot also gives rise to positive cross-correlation in
the three-lead dot.

We thank N.~J.~Craig for device fabrication and H.-A.~Engel for
valuable discussions. We acknowledge support from the NSF through
the Harvard NSEC, PHYS 01-17795, DMR-05-41988, DMR-0501796.
M.~Yamamoto and S.~Tarucha acknowledge support from the DARPA QuIST
program, the Grant-in-Aid for Scientific Research A (No. 40302799),
the MEXT IT Program and the Murata Science Foundation.


\small
\begin{thebibliography}{31}
\expandafter\ifx\csname natexlab\endcsname\relax\def\natexlab#1{#1}\fi
\expandafter\ifx\csname bibnamefont\endcsname\relax
  \def\bibnamefont#1{#1}\fi
\expandafter\ifx\csname bibfnamefont\endcsname\relax
  \def\bibfnamefont#1{#1}\fi
\expandafter\ifx\csname citenamefont\endcsname\relax
  \def\citenamefont#1{#1}\fi
\expandafter\ifx\csname url\endcsname\relax
  \def\url#1{\texttt{#1}}\fi
\expandafter\ifx\csname urlprefix\endcsname\relax\def\urlprefix{URL }\fi
\providecommand{\bibinfo}[2]{#2} \providecommand{\eprint}[2][]{\url{#2}}


\bibitem[{\citenamefont{Noise Reviews}(00)}]{blanter00-05}
\bibinfo{author}{\bibfnamefont{Ya.~M.}~\bibnamefont{Blanter}} \bibnamefont{and}
 \bibinfo{author}{\bibfnamefont{M.}~\bibnamefont{B\"{u}ttiker}},
 \bibinfo{journal}{Phys.~Rep.} \textbf{\bibinfo{volume}{336}},
 \bibinfo{pages}{1} (\bibinfo{year}{2000}).
\bibinfo{author}{\bibfnamefont{Ya.~M.}~\bibnamefont{Blanter}},
 \bibinfo{journal}{cond-mat/0511478} (\bibinfo{year}{2005}).

\bibitem[{\citenamefont{Scattering Theory for Noise}(1990)}]{buttiker90-92}
\bibinfo{author}{\bibfnamefont{M.}~\bibnamefont{B{\"{u}}ttiker}},
  \bibinfo{journal}{Phys.\ Rev.\ Lett.} \textbf{\bibinfo{volume}{65}},
  \bibinfo{pages}{2901} (\bibinfo{year}{1990});
\bibinfo{author}{\bibfnamefont{M.}~\bibnamefont{B{\"{u}}ttiker}},
  \bibinfo{journal}{Phys.\ Rev.\ B} \textbf{\bibinfo{volume}{46}},
  \bibinfo{pages}{12485} (\bibinfo{year}{1992}).

\bibitem[{\citenamefont{HBT QHall}(1999)}]{henny99}
\bibinfo{author}{\bibfnamefont{M.}~\bibnamefont{Henny}} \bibnamefont{\textit{et~al.}},
  \bibinfo{journal}{Science} \textbf{\bibinfo{volume}{284}},
  \bibinfo{pages}{296} (\bibinfo{year}{1999}).
\bibinfo{author}{\bibfnamefont{S.}~\bibnamefont{Oberholzer}}
  \bibnamefont{\textit{et~al.}}, \bibinfo{journal}{Physica E}
  \textbf{\bibinfo{volume}{6}}, \bibinfo{pages}{314} (\bibinfo{year}{2000}).

\bibitem[{\citenamefont{HBT Zero Field}(1999)}]{oliver99}
\bibinfo{author}{\bibfnamefont{W.~D.} \bibnamefont{Oliver}}
  \bibnamefont{\textit{et~al.}}, \bibinfo{journal}{Science}
  \textbf{\bibinfo{volume}{284}}, \bibinfo{pages}{299} (\bibinfo{year}{1999}).

\bibitem[{\citenamefont{Proposal of Voltage Probe Induced Positive Cross-correlation}(2000)}]{texier00}
\bibinfo{author}{\bibfnamefont{C.}~\bibnamefont{Texier}} \bibnamefont{and}
  \bibinfo{author}{\bibfnamefont{M.}~\bibnamefont{B{\"{u}}ttiker}},
  \bibinfo{journal}{Phys.\ Rev.\ B} \textbf{\bibinfo{volume}{62}},
  \bibinfo{pages}{7454} (\bibinfo{year}{2000}).

\bibitem[{\citenamefont{Reverse the Sign of Cross-correlation}(2003)}]{buttiker03}
\bibinfo{author}{\bibfnamefont{M.}~\bibnamefont{B{\"{u}}ttiker}},
\bibnamefont{in}
\bibinfo{journal}{\textit{Quantum Noise in Mesoscopic Physics}, NATO Science Series II} \textbf{\bibinfo{volume}{97}},
\bibnamefont{edited by}~\bibnamefont{Yu.~V.}~\bibnamefont{Nazarov}
(\bibinfo{year}{Kluwer, Dordrecht, 2003}),
\bibinfo{journal}{cond-mat/0209031}.

\bibitem[{\citenamefont{External Feedback Induced Positive Cross-correlation}(2005)}]{wu05}
\bibinfo{author}{\bibfnamefont{S.-T.} \bibnamefont{Wu}} \bibnamefont{and}
  \bibinfo{author}{\bibfnamefont{S.}~\bibnamefont{Yip}},
  \bibinfo{journal}{Phys.\ Rev.\ B} \textbf{\bibinfo{volume}{72}},
  \bibinfo{pages}{153101} (\bibinfo{year}{2005}).

\bibitem[{\citenamefont{Controlled Inelastic Scattering Induced Positive Cross-correlation}(2006)}]{rychkov06}
\bibinfo{author}{\bibfnamefont{V.}~\bibnamefont{Rychkov}} \bibnamefont{and}
  \bibinfo{author}{\bibfnamefont{M.}~\bibnamefont{B{\"{u}}ttiker}},
  \bibinfo{journal}{Phys.\ Rev.\ Lett.} \textbf{\bibinfo{volume}{96}},
  \bibinfo{pages}{166806} (\bibinfo{year}{2006}).

\bibitem[{\citenamefont{Cotunneling Induced Super-Poissonian Noise}(2001)}]{sukhorukov01}
\bibinfo{author}{\bibfnamefont{E.~V.} \bibnamefont{Sukhorukov}}
  \bibnamefont{\textit{et~al.}}, \bibinfo{journal}{Phys.\ Rev.\ B}
  \textbf{\bibinfo{volume}{63}}, \bibinfo{pages}{125315}
  (\bibinfo{year}{2001}).

\bibitem[{\citenamefont{Double Dot Master Equation}(2003)}]{kiesslich03}
\bibinfo{author}{\bibfnamefont{G.}~\bibnamefont{Kiesslich}}
  \bibnamefont{\textit{et~al.}}, \bibinfo{journal}{Phys.\ Rev.\ B}
  \textbf{\bibinfo{volume}{68}}, \bibinfo{pages}{125320}
  (\bibinfo{year}{2003}).

\bibitem[{\citenamefont{Dynamical Channel Blockade - Two-lead}(2005)}]{belzig05}
\bibinfo{author}{\bibfnamefont{W.}~\bibnamefont{Belzig}},
  \bibinfo{journal}{Phys.\ Rev.\ B} \textbf{\bibinfo{volume}{71}},
  \bibinfo{pages}{R161301} (\bibinfo{year}{2005}).

\bibitem[{\citenamefont{Dynamical Channel Blockade - Three-lead}(2004)}]{cottet04}
\bibinfo{author}{\bibfnamefont{A.}~\bibnamefont{Cottet}} \bibnamefont{\textit{et~al.}},
  \bibinfo{journal}{Phys.\ Rev.\ B} \textbf{\bibinfo{volume}{70}},
  \bibinfo{pages}{115315} (\bibinfo{year}{2004}).
\bibinfo{author}{\bibfnamefont{A.}~\bibnamefont{Cottet}} \bibnamefont{\textit{et~al.}},
  \bibinfo{journal}{Phys.\ Rev.\ Lett.} \textbf{\bibinfo{volume}{92}},
  \bibinfo{pages}{206801} (\bibinfo{year}{2004}).

\bibitem[{\citenamefont{Thielmann et~al.}(2001)}]{thielmann05}
\bibinfo{author}{\bibfnamefont{A.} \bibnamefont{Thielmann}}
  \bibnamefont{\textit{et~al.}}, \bibinfo{journal}{Phys.\ Rev.\ Lett.}
  \textbf{\bibinfo{volume}{95}}, \bibinfo{pages}{146806}
  (\bibinfo{year}{2005}).



\bibitem[{\citenamefont{Negative Conductance and Super-Poissoinian Noise}(1998)}]{Iannaccone98}
\bibinfo{author}{\bibfnamefont{G.}~\bibnamefont{Iannaccone}}
  \bibnamefont{\textit{et~al.}}, \bibinfo{journal}{Phys.\ Rev.\ Lett.}
  \textbf{\bibinfo{volume}{80}}, \bibinfo{pages}{1054}
  (\bibinfo{year}{1998}).

\bibitem[{\citenamefont{Safonov et~al.}(2003)}]{safonov03}
\bibinfo{author}{\bibfnamefont{S.~S.} \bibnamefont{Safonov}}
  \bibnamefont{\textit{et~al.}}, \bibinfo{journal}{Phys.\ Rev.\ Lett.}
  \textbf{\bibinfo{volume}{91}}, \bibinfo{pages}{136801}
  (\bibinfo{year}{2003}).

\bibitem[{\citenamefont{Chen and Webb}(2006)}]{chenPRB06}
\bibinfo{author}{\bibfnamefont{Y.}~\bibnamefont{Chen}} \bibnamefont{and}
  \bibinfo{author}{\bibfnamefont{R.~A.} \bibnamefont{Webb}},
  \bibinfo{journal}{Phys.\ Rev.\ B} \textbf{\bibinfo{volume}{73}},
  \bibinfo{pages}{35424} (\bibinfo{year}{2006}).

\bibitem[{\citenamefont{Barthold et~al.}(2006)}]{barthold06}
\bibinfo{author}{\bibfnamefont{P.} \bibnamefont{Barthold}}
  \bibnamefont{\textit{et~al.}}, \bibinfo{journal}{Phys.\ Rev.\ Lett.}
  \textbf{\bibinfo{volume}{96}}, \bibinfo{pages}{246804}
  (\bibinfo{year}{2006}).

\bibitem[{\citenamefont{Chen and Webb}(2006)}]{chenPRL06}
\bibinfo{author}{\bibfnamefont{Y.}~\bibnamefont{Chen}} \bibnamefont{and}
  \bibinfo{author}{\bibfnamefont{R.~A.} \bibnamefont{Webb}},
  \bibinfo{journal}{Phys.\ Rev.\ Lett.} \textbf{\bibinfo{volume}{97}},
  \bibinfo{pages}{66604} (\bibinfo{year}{2006}).

\bibitem[{\citenamefont{Oberholzer et~al.}(2006)}]{oberholzer06}
\bibinfo{author}{\bibfnamefont{S.}~\bibnamefont{Oberholzer}}
  \bibnamefont{\textit{et~al.}}, \bibinfo{journal}{Phys.\ Rev.\ Lett.}
  \textbf{\bibinfo{volume}{96}}, \bibinfo{pages}{46804}
  (\bibinfo{year}{2006}).

\bibitem[{\citenamefont{Onac et~al.}(2006)}]{onac06}
\bibinfo{author}{\bibfnamefont{E.}~\bibnamefont{Onac}} \bibnamefont{\textit{et~al.}},
  \bibinfo{journal}{Phys.\ Rev.\ Lett.} \textbf{\bibinfo{volume}{96}},
  \bibinfo{pages}{26803} (\bibinfo{year}{2006}).


\bibitem[{\citenamefont{Gustavsson et~al.}()}]{gustavsson06}
\bibinfo{author}{\bibfnamefont{S.}~\bibnamefont{Gustavsson}} \bibnamefont{\textit{et~al.}},
  \bibinfo{journal}{Phys.\ Rev.\ B} \textbf{\bibinfo{volume}{74}},
  \bibinfo{pages}{195305} (\bibinfo{year}{2006}).

\bibitem[{\citenamefont{Zarchin et~al.}(2006)}]{Zarchin06}
\bibinfo{author}{\bibfnamefont{O.}~\bibnamefont{Zarchin}} \bibnamefont{\textit{et~al.}},
  \bibinfo{journal}{Phys.\ Rev.\ Lett.} \textbf{\bibinfo{volume}{98}},
  \bibinfo{pages}{66801} (\bibinfo{year}{2007}).

\bibitem[{\citenamefont{McClure et~al.}(2006)}]{McClure06}
\bibinfo{author}{\bibfnamefont{D.~T.}~\bibnamefont{McClure}} \bibnamefont{\textit{et~al.}},
  \bibinfo{journal}{Phys.\ Rev.\ Lett.} \textbf{\bibinfo{volume}{98}},
  \bibinfo{pages}{56801} (\bibinfo{year}{2007}).

\bibitem[{\citenamefont{Eto~et~al.}(1997)}]{eto97}
\bibinfo{author}{\bibfnamefont{M.}~\bibnamefont{Eto}}
  \bibnamefont{\textit{et~al.}}, \bibinfo{journal}{Jpn.\ J.\ Appl.\ Phys.}
  \textbf{\bibinfo{volume}{36}}, \bibinfo{pages}{4004}
  (\bibinfo{year}{1997}).


\bibitem[{\citenamefont{DiCarlo et~al.}(2006)}]{techniques}
\bibinfo{author}{\bibfnamefont{L.} \bibnamefont{DiCarlo}}
  \bibnamefont{\textit{et~al.}}, \bibinfo{journal}{Rev.\ Sci.\ Inst.}
  \textbf{\bibinfo{volume}{77}}, \bibinfo{pages}{73906}
  (\bibinfo{year}{2006}).



\bibitem[{\citenamefont{ExtractSi}()}]{ExtractSi}
The dot's intrinsic current noises are extracted by solving the
Langevin equations for finite-impedance external
circuits~\cite{blanter00-05}:
\begin{eqnarray*}
\Sione & = & a_{11}^2 S_{V1} + a_{21}^2 S_{V2} + 2a_{11}
a_{21}S_{V12} -
4\kb\te/R \\
\Sitwo & = & a_{12}^2 S_{V1} + a_{22}^2 S_{V2} + 2a_{12}
a_{22}S_{V12} - 4\kb\te/R \\
\Six & = & a_{11}a_{12}S_{V1} + a_{21}a_{22}S_{V2} +
(a_{11}a_{22}+a_{12}a_{21})S_{V12},
\end{eqnarray*}
where $a_{11(22)} = 1/R - g_{11(22)}$, $a_{12(21)} = -g_{12(21)}$
and $R$ is the RLC resonator parallel resistance.


\bibitem[{\citenamefont{Hansres Thesis}(2003)}]{HansresThesis}
\bibinfo{author}{\bibfnamefont{H.-A.}~\bibnamefont{Engel}},
 \bibinfo{journal}{Ph.D. thesis, University of Basel} (\bibinfo{year}{2003}).

\bibitem[{\citenamefont{Master Equation for Noise}(1993)}]{hershfield93}
\bibinfo{author}{\bibfnamefont{S.}~\bibnamefont{Hershfield}}
  \bibnamefont{\textit{et~al.}}, \bibinfo{journal}{Phys.\ Rev.\ B}
  \textbf{\bibinfo{volume}{47}}, \bibinfo{pages}{1967} (\bibinfo{year}{1993}).


\bibitem[{\citenamefont{Explain M.L. model details}()}]{ExplainMLmodel}
For computational reasons, we limit the calculation to $n=m=3$. For
simplicity, we assume equal level spacings, symmetric tunnel
barriers, and an exponential dependence of the tunneling rates on
level energy: $\Delta\varepsilon^l \equiv \varepsilon^l_d -
\varepsilon^0_d = l\times\delta$ and $\gamma^l_{0} = \gamma^l_{1} =
\Gamma \exp(\kappa \Delta\varepsilon^l)$, where $l=-3,...,0,...,3$
is the level index, $\varepsilon^l_d$ is the energy of
level~$l$, and $\gamma^l_{0(1)}$ is the tunneling rate from level~$l$ to
reservoir~0(1). We chose $\delta = 150~\mathrm{\mu eV}$,
$\Gamma = 15~\mathrm{GHz}$ and $\kappa = 0.001~\mathrm{(\mu
eV)^{-1}}$ to fit the data in Figs.~1(d) and 1(e).

\bibitem[{\citenamefont{Explain the Difference between g12 and g21}()}]{ExplainG12andG21}
At zero bias, the fluctuation-dissipation theorem requires $\Six = - 2
\kb \te (\gonetwo + \gtwoone)$, but $\gonetwo = \gtwoone$ at zero
bias and zero magnetic field. However, the fact that $\gonetwo \neq
\gtwoone$ at finite bias, observed experimentally, requires
knowledge of the full conductance matrix to properly extract
$S_{1,2}$ and $\Six$~\cite{ExtractSi}.


\end{thebibliography}
\end{document}